\documentclass[twocolumn,twoside,slac_two]{revtex4}
\usepackage{graphicx}
\usepackage{fancyhdr}
\begin{document}

\title{Lepton Flavour Violation and \boldmath$B_s$ Leptonic Final States at the LHC}

%

\author{P. \v Rezn\' i\v cek, for the ATLAS and CMS collaborations}
\affiliation{IPNP, Faculty of Mathematics and Physics, Charles University in Prague, Czech Republic}

\begin{abstract}

An~overview of~ATLAS and~CMS studies of~$B_s$ leptonic decays providing constraints on~the~lepton flavour violation phenomenon is presented.
Except direct lepton flavour violating final states $B_s \to l_1^+ l_2^-$, constraints can also be set by~a~measurement of~$B_s \to \mu^+ \mu^-$ decay, whose branching ratio in~some theoretical models correlates with~a~branching ratio of~$B_s \to l_1^+ l_2^-$, $\tau \to \mu \mu \mu$ and~some other lepton flavour violating decays.
In this paper, the~feasibility of~measurements of~$B_s \to \mu^+ \mu^-$ decay is described, including the~present status, the~trigger and~the~offline analysis strategies and~the~expected reach in~the~branching ratio measurement.
The~ATLAS and~CMS experiments foresee to~provide $3\sigma$ evidence of~Standard Model $B_s \to \mu^+ \mu^-$ branching ratio by~the~end of~LHC low-luminosity stage ($30\;\mathrm{fb}^{-1}$).
Also a~CMS study of~the~$\tau \to \mu \mu \mu$ decay and~an~initial particle-level based study of~the~$B_s \to \tau \mu$ decay are presented.
A~sensitivity of~$\sim 10^{-8}$ for~the~$\tau \to \mu \mu \mu$ branching ratio measurement is predicted by~CMS.

\end{abstract}

\maketitle

\thispagestyle{fancy}


\section{\label{sec:Intro} Introduction}

The~flavour mixing is a~common phenomenon in~the~quark sector as well as mixing by~means of~neutrino oscillations in~the~lepton sector.
But in~the~Standard Model (SM) there is no such phenomenon in~connection with~charged leptons.
It is a~built-in property based on~experimental results, without a~corresponding symmetry in~the~SM.
With~massive neutrinos, charged lepton flavour violation (LFV) is possible, but~the~cross-section would be negligible due to~the~smallness of~the~neutrino mass and~the~GIM cancellation mechanism.
Thus the~branching ratio (BR) would not be detectable by~the~present experiments.
However, beyond the~SM there are theoretical models allowing enhanced LFV in~the~charged leptons sector too.

Due to~high LHC luminosity ($10^{33}/10^{34}\;\mathrm{cm}^{-2}\mathrm{s}^{-1}$ in low/high luminosity stage), the~ATLAS and~CMS experiments have excellent capabilities of~low-BR decay measurements.
One sector, in~which~LFV decays can be searched for, are the~purely leptonic $B_s$ decays.
Except for~an~initial particle-level based test of~the~feasibility of~the~measurement of~$B_s\to\tau\mu$ at~a~general purpose detector for~a collider experiment~\cite{Langenegger:2008zz} (see section~\ref{sec:Bstaumu}), there are no ATLAS/CMS studies concerning direct LFV in~leptonic $B_s$ decays yet.
However, the~phenomenon can also be studied indirectly using correlations between LFV and non-LFV decays.
Examples of~such correlations can be found e.g. in~constrained or~general flavour-universal MSSM models~\cite{Dedes:2002rh}, relating the~BR of~a~very rare $B_s\to\mu^+\mu^-$ decay to~the~LFV decays $B_s\to l_1^+ l_2^-$ and~$\tau\to\mu\mu\mu$.
Both the~ATLAS and~the~CMS experiments performed detailed studies of~a~possible measurement of~flavour-changing neutral currents (FCNC) in~$B_s\to\mu^+\mu^-$ decays~\cite{CSCbphys:2008,Langenegger:2006jk}.
The~trigger and~the~analysis strategy, the~background sources, the~expected performance and~the~reach in~this channel are described in~section~\ref{sec:Bsmumu}.
CMS also studied $\tau\to\mu\mu\mu$ LFV decays~\cite{Santinelli:2002cms}.
The~analysis details and~resulting performance are presented in~section~\ref{sec:taumumumu}.

\section{\label{sec:Bsmumu} \boldmath$B_s\to\mu^+\mu^-$ at~ATLAS and~CMS}

The~di-muonic FCNC B-decays $B_s\to\mu^+\mu^-$ can occur in~the~SM through higher order diagrams only and~are also helicity suppressed, which results in~the~prediction of~a~very low BR of~$(3.35\pm 0.32)\cdot 10^{-9}$~\cite{Blanke:2006ig}.
This small BR provides room for~New Physics (NP) effects, which may enhance or~suppress BR significantly.
Present best limits are set by~Tevatron: a~CDF measurement at~$2\;\mathrm{fb}^{-1}$ sets the~upper limit on~the~BR to~$5.8\cdot 10^{-8}$ at $95\%\;\mathrm{C.L.}$~\cite{Aaltonen:2007kv}.
The~projected improvement of~the~Tevatron results~\cite{Krutelyov:hcp2006} does not~foresee reaching the~SM BR.
In~constrast the~LHC experiments will even be~able to~measure a~suppressed BR w.r.t to~the~SM prediction, especially after entering the~high luminosity stage of~LHC.

\subsection{Trigger Strategies}

Both the~ATLAS and~CMS triggers for~$B_s\to\mu^+\mu^-$~\cite{Buchalla:2008jp} are at~the~first level (L1) based on~a~detection of~two high-$p_\mathrm{T}$ muons and~at~the~high level trigger (HLT) the~$B_s$ vertex is fitted and~cuts on~its quality, eventually position and~di-muon invariant mass are applied.
Presently, the~ATLAS $B_s\to\mu^+\mu^-$ study requires two $p_\mathrm{T}>6\;\mathrm{GeV}$ muons at~L1, while the~CMS study lets to~pass through muons with~$p_\mathrm{T}>3\;\mathrm{GeV}$.
However, based on~the~first data, the~cuts are subject to~change with~the~luminosity in~order to~keep an~acceptable~trigger-output rate.
At~the~luminosity of~$10^{33}\;\mathrm{cm}^{-2}\mathrm{s}^{-1}$ the~ATLAS L1 rate is expected to~be $360\;\mathrm{Hz}$ and~the~CMS L1 rate $450\;\mathrm{Hz}$.
The~detectors have similar pseudorapidity coverage $|\eta|<2.5$.

At~the~ATLAS HLT, firstly the~muons' detection is confirmed by~information from~precision chambers and~calorimeters.
Tracks reconstruction is then performed in~a region around the~two L1 muon directions and~di-muon candidates are defined.
Applying a~vertex fit procedure a~quality cut $\chi^2 < 10$ and~a loose invariant mass cut $M_{\mu^+\mu^-} < 7\;\mathrm{GeV}$ are currently required.
However, these~cuts could be tightened or~new cuts could be introduced (e.g. a~cut on~$B_s$ transverse decay length $L_{xy}$) when a~too high output rate is experienced with~first LHC data.
The~overall L1 and~HLT efficiency of~46\% was found on~a sample of~simulated signal events with~muons $p_\mathrm{T} > 6\;\mathrm{GeV}$ and~$|\eta| < 2.5$.
At~a~very low starting luminosity of~$\sim 10^{31}\;\mathrm{cm}^{-2}\mathrm{s}^{-1}$ the~L1 threshold cut on~the~muons' $p_\mathrm{T}$ can be lowered down to~$4\;\mathrm{GeV}$ and~the~track reconstruction could be performed in~the~full volume of~the~ATLAS inner detector.

The~CMS HLT firstly verifies L1 muons, then the~primary vertex is searched (only the~3 most-significant vertices are considered) and~tracks of~$p_\mathrm{T} > 4\;\mathrm{GeV}$ are reconstructed in~cones around the~muon directions.
The~tracks are always formed from~less than six detector hits in~order to~speed up the~reconstruction process, but at~the~expense of~a~certain performance degradation ($74\;\mathrm{MeV}$ $B_s$ mass resolution to be compared to~$32\;\mathrm{MeV}$ in~the~offline analysis).
After vertex fitting, a~decay length cut of~$L_{3D}\ge 150\;\mu\mathrm{m}$, a~vertex quality cut of~$\chi^2\le 20$ and~a~$B_s$ invariant mass window of~$150\;\mathrm{MeV}$ width are applied.
The~output rate after the~HLT is expected to be below $1.7\;\mathrm{Hz}$.

\subsection{Offline Analysis}

Due to~the~clear experimental signature, the~selection criteria are limited to~the~following cuts on: the~$B_s$ invariant mass window, the~secondary vertex displacement and the~vertex-fit quality, the~$B_s$ momentum pointing to~the~primary vertex, the~isolation of~the~muon pair and the di-muon opening angle.
The~detector-driven mass resolution should either be sufficient to~distinguish $B_s$ from~$B_d$, or a~joint analysis of~both decay channels has to~be performed.
The~vertex position and~fit-quality cuts reduce combinatorial background originating from~the~primary vertex.
The~$B_s$-pointing-to-the-primary-vertex requirement suppresses events with~particles invisible to~the~detector and~originating at~the~di-muon vertex.
Finally, the~isolation constraint helps to~similarly suppress visible particles originating from~the~di-muon vertex.

\subsubsection{ATLAS}

In~the~ATLAS offline analysis high-$p_\mathrm{T}$ muon tracks are combined to~di-muon pairs and~the~appropriate vertex is fitted.
A~preselection of~the~$B_s$ candidates is performed by~requiring $4\;\mathrm{GeV} < M_{\mu^+\mu^-} < 7.3\;\mathrm{GeV}$ for~the~di-muon pair invariant mass, vertex fit quality $\chi^2 < 10$ and~a~transverse decay length $L_{xy} < 20\;\mathrm{mm}$.
A~final asymmetric invariant mass cut of~$M_{B_s}-\sigma_{B_s} < M_{\mu^+\mu^-} < M_{B_s}+2\sigma_{B_s}$ is used in~order to~separate $B_s$ candidates from~$B_d$ background.
The~di-muonic invariant mass resolution at~ATLAS is $70\;\mathrm{MeV}$ for~the~barrel region ($\eta_{\mu^+\mu^-} < 1.1$) and~$124\;\mathrm{MeV}$ for~the~end-cap.
A~combined value $\sigma_{B_s} = 90\;\mathrm{MeV}$ is estimated as the~invariant mass resolution for~the~signal events.
Secondary vertex separation is defined by~requiring a~$B_s$ decay length $L_\mathrm{xy} > 0.5\;\mathrm{mm}$.
In~order to~assure the~$B_s$ pointing to~the~primary vertex, the~$B_s$ momentum and~the~position vector from~the~primary to~the~secondary vertex have to~be parallel within an~angle of~$1^\circ$.
The~di-muon pair isolation is defined using $I_{\mu^+\mu^-} = p_\mathrm{T}(B_s)/(p_\mathrm{T}(B_s)+\sum_{tracks}|p_\mathrm{T}|) > 0.9$ condition, where the~sum runs over all tracks of~$p_\mathrm{T} > 1\;\mathrm{GeV}$ (excluding signal muons) reconstructed within a~cone~of~$\Delta R = \sqrt{\Delta\phi^2+\Delta\eta^2} < 1$ around the~di-muon momentum.

\subsubsection{CMS}

The~CMS offline analysis is similar to~the~ATLAS one.
Due to~a~better $B_s$ invariant mass resolution of~$32\;\mathrm{MeV}$, a symmetric mass window of~$M_{B_s}-100\;\mathrm{MeV} < M_{\mu^+\mu^-} < M_{B_s}+100\;\mathrm{MeV}$ is used.
A~decay length significance $L_\mathrm{xy}/\sigma_\mathrm{xy} \ge 18$ and~vertex-fit $\chi^2 \le 1$ cuts are applied.
The~$B_s$ momentum and~the~vector from~the~primary to~the~secondary vertex are required to be aligned with~each other better than $5.7^\circ$.
The~muon pair isolation criterion uses a~$I_{\mu^+\mu^-} \ge 0.85$ cut, with~the~$I_{\mu^+\mu^-}$ defined similarly as~in~ATLAS analysis, only applying lower $p_\mathrm{T}$ cut of~$p_\mathrm{T} > 0.9\;\mathrm{MeV}$ on~the~tracks considered in~the~cone around the~di-muon momentum.
Lastly, the~opening angle $\Delta R(\mu^+,\mu^-)$ between the~two muon tracks is required to satisfy $0.3 < \Delta R(\mu^+,\mu^-) < 1.2$.

\subsection{\label{sec:BmumuBackrounds} Background Composition}

The~background comes from~random combinatorics of~high-$p_\mathrm{T}$ muons, especially in~events with~$b\bar{b}$ pairs, where the~eventual di-muon vertex position is naturally displaced from~the~primary vertex by~a~distance typical for~$B$-hadrons.
The~two muon candidates can~come from~semileptonic decays of~$b$ and~$\bar{b}$ quarks or~from a~cascade decay of~either the~$b$ or~$\bar{b}$ quark.
The~extremely low BR of~the~signal decay channel allows that also rare processes (often not included in standard MC generators) can significantly contribute to~the~background, as do misidentified hadron-tracks.
These background contributions can be evaluated by~a MC study of~dedicated exclusive background channels.
However, the~inclusive $b\bar{b} \to \mu^+\mu^-X$ background event generation and~the~corresponding full detector simulation are limited by~the~available CPU resources.
Thus, events need to be filtered already at~the~generator level and~a~factorization of~the~analysis cuts and~a~study of~correlations between that cuts need to be performed.
The~limited resources still lead to~a~very limited precision of~the~estimated number of~background events, and as such are awaiting improvement by~the~first LHC data.
The~ATLAS and~CMS expectation of~the~inclusive background at~an~integrated luminosity of~$L_\mathrm{int} = 10\;\mathrm{fb}^{-1}$ yields $14^{+13}_{-10}$ and~$13.8^{+22.0}_{-13.8}$ events respectively.

The~background caused by~rare decays or~misidentification of~muons can be divided into~four groups depending on~the~mechanism mimicking the~signal events.
Given that the~false $B_s$ signals come from $B$-hadron decays with~two oppositely charged high-$p_\mathrm{T}$ tracks (real muons or~misidentified hadrons) and~remaining soft particles in~the~final state (not seen), the~resulting di-muon invariant mass could be close to~the~$B$-hadron mass.
Therefore the~mass resolution is a~crucial factor when~rejecting these backgrounds.
The~signal and~selected backgrounds invariant mass distributions corresponding to~a~measurement at~ATLAS after one~year of~low luminosity running is shown at~Figure~\ref{fig:BackgroundMisId}.
\begin{itemize}

\item
The~first class of~background is formed by~rare $B$-decays that contain three muons coming from~the~same (or~a~negligibly close) vertex.
A~fake $B$-signal can appear when one of~the~muons and~the~remaining neutrino are soft and~thus not~being considered in~the~signal decay reconstruction.
Representatives of~this background class are:
\\
\begin{tabular}{ll}
$\;\;\;\;B^+   \to        \mu^+\mu^- \mu^+\nu_{\mu}$ & ($\mathrm{BR} \sim 5   \cdot 10^{-6}$),    \\
$\;\;\;\;B_c^+ \to        \mu^+\mu^- \mu^+\nu_{\mu}$ & ($\mathrm{BR} \sim 5   \cdot 10^{-6}$) and \\
$\;\;\;\;B_c^+ \to J/\psi(\mu^+\mu^-)\mu^+\nu_{\mu}$ & ($\mathrm{BR} \sim 1.2 \cdot 10^{-3}$).    \\
\end{tabular}
\\
\\
Both the~ATLAS study of~the~first two and~the~CMS study of~all three background decays found them to~be negligible.

\item
Another class of~rare-decay contributions to~the~background is composed of~$B$-hadron decays to~two muons and~a~soft photon or~pion:
\\
\begin{tabular}{ll}
$\;\;\;\;B_{(s)} \to \mu^+ \mu^- \gamma$ & ($\mathrm{BR} \sim 2 \cdot 10^{-8}$),    \\
$\;\;\;\;B       \to \mu^+ \mu^- \pi^0$  & ($\mathrm{BR} \sim 2 \cdot 10^{-8}$) and \\
$\;\;\;\;B^+     \to \mu^+ \mu^- \pi^+$  & ($\mathrm{BR} \sim 2 \cdot 10^{-8}$).    \\
\end{tabular}
\\
\\
CMS tested the~first and~ATLAS all~three possible background decays resulting in~negligible contributions.

\item
Although the~misidentification of~hadron tracks as muons is a~rare effect (typical probability of~the~order of~$0.1\%$) it can cause a~dangerous background for~the~very rare decays.
A~study of~$B$-hadron decays to~a~muon, hadron and~a~soft neutrino was performed, accounting for~the~hadron track misidentification in~the~following decays:
\\
\begin{tabular}{ll}
$\;\;\;\;B   \to \pi^- \mu^+ \nu_{\mu}$ & ($\mathrm{BR} \sim 1.4 \cdot 10^{-4}$) and \\ 
$\;\;\;\;B_s \to K^-   \mu^+ \nu_{\mu}$ & ($\mathrm{BR} \sim 1.4 \cdot 10^{-4}$).    \\ 
\end{tabular}
\\
As~in~the~previous cases the studies showed an~insignificant rate of~the~fake signal decays.

\item
An~isolated $B_s\to\mu^+\mu^-$ fake candidate can also be formed when a~double misidentification of~the~hadrons as muons occurs in~two body $B$-hadron decays:
\\
\begin{tabular}{ll}
$\;\;\;\;B_s       \to K^+    K^-$   & ($\mathrm{BR} \sim 2.4 \cdot 10^{-5}$),     \\ 
$\;\;\;\;B_s       \to \pi^+  \pi^-$ & ($\mathrm{BR} \sim 0.4 \cdot 10^{-6}$),     \\ 
$\;\;\;\;B_s       \to \pi^+  K^-$   & ($\mathrm{BR} \sim 5.0 \cdot 10^{-6}$),     \\ 
$\;\;\;\;B         \to \pi^+  \pi^-$ & ($\mathrm{BR} \sim 5.1 \cdot 10^{-6}$),     \\ 
$\;\;\;\;B         \to K^+    \pi^-$ & ($\mathrm{BR} \sim 2.0 \cdot 10^{-5}$),     \\ 
$\;\;\;\;\Lambda_b \to \rho^+ \pi^-$ & ($\mathrm{BR} \sim 1   \cdot 10^{-6}$), and \\ 
$\;\;\;\;\Lambda_b \to \rho^+ K^-$   & ($\mathrm{BR} \sim 2   \cdot 10^{-6}$).     \\ 
\end{tabular}
\\
\\
The~ATLAS study found a~negligible contribution to~the~background from~the~first three decays in~the~list.
Similarly in~the~CMS study less than 0.3 fake signal decays for~each of~the~decays listed above were expected for~$L_\mathrm{int} = 10\;\mathrm{fb}^{-1}$.

\begin{figure}[h]
\centering
\includegraphics[width=72mm,height=62mm]{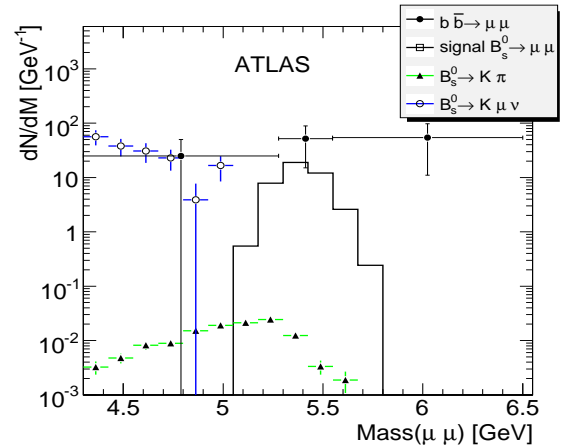}
\caption{Di-muonic invariant mass spectrum of~the~signal $B$-hadrons, inclusive $b\bar{b}$ contribution and~selected misidentification-based background channels~\cite{CSCbphys:2008}. Statistics corresponds to~$L_\mathrm{int} = 10\;\mathrm{fb}^{-1}$.}
\label{fig:BackgroundMisId}
\end{figure}

\end{itemize}

\subsection{Expected Reach}

The~performed studies predict a~$3\sigma$ evidence of~the~SM BR already by~the~end of~the~initial LHC low-luminosity stage ($L_\mathrm{int} = 30\;\mathrm{fb}^{-1}$).
The~observation at~$5\sigma$ level will be~possible after the~first year of~LHC running at~high luminosity, provided the~background will be at~the~level expected from~extrapolations.
The~BR upper limits can be calculated using the~Bayesian approach~\cite{Hebbeker:2001}.
For~example, picking up the~CMS expectation of~the~number of~signal events (supposing SM BR): $6.1\pm0.6_{stat}\pm1.5_{sys}$, and~accounting also for~the~background events expected as mentioned in~section \ref{sec:BmumuBackrounds}, the~upper $90\%\;\mathrm{C.L.}$ limit would be set to~$1.4\cdot 10^{-8}$.

\begin{table}[h]
\begin{center}
\caption{Expected $B_s\to\mu^+\mu^-$ signal yield at~given integrated luminosity. The~ATLAS results for~$100\;\mathrm{fb}^{-1}$ are given for~older studies~\cite{Ball:2000ba}$^*$.}
\begin{tabular}{|l|c|c|c|c|c|}
\hline
Experiment & $2\;\mathrm{fb}^{-1}$ & $10\;\mathrm{fb}^{-1}$ & $30\;\mathrm{fb}^{-1}$ & $100\;\mathrm{fb}^{-1}$ & $130\;\mathrm{fb}^{-1}$ \\
\hline
ATLAS      & 1.1 & 5.7 & 17.1 & 92$^*$ & 109$^*$ \\
CMS        & 1.2 & 6.1 & 18.3 &        &         \\
\hline
\end{tabular}
\label{tab:BsmumuSignalEvents}
\end{center}
\end{table}

In~Table~\ref{tab:BsmumuSignalEvents} the~number of~signal events is summarized at~the~following stages: after approximately the~first 2 months ($2\;\mathrm{fb}^{-1}$) and the~first year of~LHC running ($10\;\mathrm{fb}^{-1}$), at~the~end of~low luminosity stage ($30\;\mathrm{fb}^{-1}$), yield of~a~year of~LHC high luminosity ($100\;\mathrm{fb}^{-1}$) and after four years of~running ($130\;\mathrm{fb}^{-1}$, end of~the~first year of~high luminosity).
However, there is a~theoretical uncertainty of~factor of~two in~the~$b$-production cross-section at~LHC energies and~consequently all the~mentioned number of~events scale accordingly.

\section{\label{sec:taumumumu} \boldmath$\tau\to\mu\mu\mu$ at~CMS}

The~neutrino-less $\tau$ decay BR is correlated to~the~BR of~very rare $B_s$ decays, as mentioned in~section~\ref{sec:Intro}.
Out of~the~number of~LFV $\tau$-decays, the~channel $\tau \to \mu\mu\mu$ is likely to~be the~only detectable at~ATLAS and~CMS, although studies of~the~other channels are also under consideration.
CMS has performed a~detailed study of~this particular $\tau$ decay channel~\cite{Santinelli:2002cms}, while at~ATLAS the~study is ongoing.
Beyond the~SM there is number of~theoretical possibilities allowing LFV.
A~BR($\tau \to \mu\mu\mu$) estimate within the~expected CMS sensitivity of~$10^{-8}$ is found e.g. in~mSUGRA theories with~right handed neutrinos and~universal input parameters at~the~GUT scale.
The~present $90\%\;\mathrm{C.L.}$ upper limit on~the~BR($\tau \to \mu\mu\mu$) is $3.2\cdot 10^{-8}$~\cite{Amsler:2008zz}.

The~total inclusive $\tau$ production cross-section was found to~be $\sigma(pp\to\tau+X)\sim 120\;\mu\mathrm{b}$~\cite{Santinelli:2002cms}, resulting in~$10^{12}$ $\tau$ leptons within the~CMS tracker acceptance for~$L_\mathrm{int}=10\;\mathrm{fb}^{-1}$.
There are several possible sources of~$\tau$ leptons at~the~LHC: the~$D$-mesons, $B$-mesons and~decays of~$W$ and~$Z$ bosons.
The~expected number of~$\tau$ leptons from~each of~the~sources is presented in~Table~\ref{tab:tausources}.
Although the~$W$ and~$Z$ bosons as sources of~the~$\tau$ leptons have a~much lower production cross-section than the~$D$- and~$B$-mesons, they become the~most important sources for~the~selected signal due to~the~higher $p_\mathrm{T}$ of~the~final state muons.

\begin{table}[h]
\begin{center}
\caption{Sources of~$\tau$ leptons at~the~LHC.}
\begin{tabular}{|c|c|}
\hline
Decay channel          & $N_\tau (L_\mathrm{int}=10\;\mathrm{fb}^{-1}$) \\
\hline
$W    \to\tau\nu_\tau$ & $1.7\cdot 10^{8}$  \\
$Z    \to\tau\tau$     & $3.2\cdot 10^{7}$  \\
$B^0  \to\tau X$       & $4.0\cdot 10^{11}$ \\
$B^\pm\to\tau X$       & $3.8\cdot 10^{11}$ \\
$B_s  \to\tau X$       & $7.9\cdot 10^{10}$ \\
$D_s  \to\tau X$       & $1.5\cdot 10^{12}$ \\
\hline
\end{tabular}
\label{tab:tausources}
\end{center}
\end{table}

\subsection{Trigger Strategy}

The~trigger selection of~the~$\tau \to \mu\mu\mu$ decays is based on~the~detection of~high-$p_\mathrm{T}$ muons.
At~the~first level (L1) a~single $p_\mathrm{T}>14\;\mathrm{GeV}$ muon will be required or~a~di-muon with $p_\mathrm{T} > 3\;\mathrm{GeV}$ for~each muon.
The~cuts are raised at~HLT when $p_\mathrm{T}>19\;\mathrm{GeV}$ for~single muons and~$p_\mathrm{T}>7\;\mathrm{GeV}$ for~di-muons is applied.
However, the~trigger condition will have to~be more stringent at~the~high luminosity stage due to~a~growing pile-up contribution.

\subsection{Offline Analysis}

The~offline analysis selections vary with~the~sources of~the $\tau$ leptons, but some parts are common to~all of~them.
Similarly to~the~very rare decays of~the~$B_s$ the~experimental signature is very simple.
The~basic identification requires only three good muon candidates from~the~tracker with~$p_\mathrm{T} > 4\;\mathrm{GeV}$ in~the~barrel or~with~$p_\mathrm{T} > 2.5\;\mathrm{GeV}$ in~the~endcap.
An~isolation criterion is applied for~the~case of~$\tau$ leptons coming from~the~$W$- or~$Z$ boson decays.
The~corresponding condition is to~have no charged tracks of~$p_\mathrm{T} > 0.7\;\mathrm{GeV}$ inside a cone of~$\Delta R = 0.4$ around the~muonic tracks.
Also the~reconstructed $\tau$ candidates need to~have an~invariant mass within a~window of~$M_\tau-25\;\mathrm{MeV} < M_{\mu\mu\mu} < M_\tau+25\;\mathrm{MeV}$.
The~predicted $\tau$-mass resolution is $24\;\mathrm{MeV}$.

The~main sources of~muons being able to~pass through the~trigger and~the~identification cuts are muons from~$b\bar{b}$ and~$c\bar{c}$ decays.
There are two topologically different ways to obtain three muons in~the~final state of~those decays.
In~the~first case ($\mathrm{BR}\sim10^{-3}$), one muon comes from~a~semileptonic decay of~a~heavy quark and~the~other two muons from~a~sequential decay of~the~opposite side $B$- or $D$-meson.
These events are rejected by~constraining the~maximum opening angle of~each pair of~muons.
In~the~second case ($\mathrm{BR}\sim10^{-5}$) all the~three muons come from~the~decay chain of~the~same $b$ or~$c$ quark.
The~dominant source of~background is due to~the~following decay channels:
\begin{itemize}
\item $c\bar{c} \to DxD_s \to \mu\nu_\mu\phi(\mu\mu)\;+\;X$
\item $b\bar{b} \to BxB_s \to \mu\nu_\mu D_s(K,\pi,\rho+\phi(\mu\mu))\;+\;X$
\end{itemize}
In~order to~exclude these events, every muon pair is required to~have an~invariant mass of~more than $25\;\mathrm{MeV}$ away from~the~$\phi$ nominal mass.
The~background from~$b\bar{b}$ is naturally suppressed by~the~triple muon effective mass that is shifted to higher values of~4 to 5$\;\mathrm{GeV}$.
For~the~$b\bar{b}$ events one muon is originating at~the~$B$-meson decay point and~the~other two at~the~$\phi$ decay point.
The~distance between these two points is significant and~thus the~quality of~the~three muons vertex fit or~the~maximum distance between each muon pair vertices could be significantly different from~the~signal events, thus providing another discriminating variable.

\subsubsection{$W$ bosons as source}

The~$W\to\tau\nu_\tau$ boson decay is characterized by~a~large missing transverse energy $E_\mathrm{Tmiss}$.
An~application of~the~condition~$E_\mathrm{Tmiss} > 20\;\mathrm{GeV}$ rejects the~background.
It also separates~the~signal from~$Z\to\tau\tau$, which is selected using different~criteria (see next section).

\subsubsection{$Z$ bosons as source}

With~the~$\tau$ lepton from~the~$Z$ boson source, the~presence of~the~second high-$p_\mathrm{T}$ $\tau$ is the~determining attribute.
The~second $\tau$ lepton is allowed to decay to~any~mode and~thus will be detected as~a~one or~three collimated charged tracks ($\tau$-jet), well isolated from~the~other tracks in~the~event: these~tracks are required to~fit in~a~cone of~$\Delta R = 0.03$ and~no~other $p_\mathrm{T} > 1.5\;\mathrm{GeV}$ tracks should be found within a~complementary external cone of~$\Delta R = 0.4$.
Another characteristic is the~three muons' momentum $p_{T}^{3\mu}$ that is, on~average, higher for~signal than for~the~background.
An~optimization of~the~cut leads to~the~condition $p_{T}^{3\mu} > 23\;\mathrm{GeV}$.
Finally, the~missing energy in~these events should be consistent with~the~energy of~a~$\nu_\tau$ from~the~$\tau$-jet.
Reconstructing the~invariant mass $M_Z$ from~the~three muons, the~$\tau$-jet and~$E_\mathrm{Tmiss}$, a~distribution that~peaks at~the~$Z$ boson mass is obtained (when applied to~signal events).
A~cut $M_Z > 70\;\mathrm{GeV}$ removes about 70\% of~background while keeping the~signal event loss acceptably low.

\subsubsection{$B$-mesons as source}

Due to~the~low muon $p_\mathrm{T}$, the~detection of~$\tau$ leptons coming from~a~$B$-meson decay would only be possible at~the~low luminosity stage, when muonic trigger cuts can be lower while keeping an~acceptable trigger output rate.
In~order to~select these events, a~$b$-tagging algorithm is used.
Any event containing a~jet with~three muons and~having three or~more tracks with~an~impact parameter significance greater than 2 is accepted.
The~same cut is also applied on~the~second $b$-jet in~the~event.
In~this analysis the~muonic tracks' isolation condition has to~be omitted.
The~combination of~the~$b$-tag for~the~two jets gives a~20\% efficiency for~signal events and~about 7\% efficiency for~the~$c\bar{c}$ background.

\subsubsection{$D$-mesons as source}

Although the~$\tau$-production cross-section from~$D$-mesons is the~largest one from~all the~possible sources, the~very low muon $p_\mathrm{T}$ leads to~a~suppression of~the~number of~detectable $\tau$ decays well below the~contribution of~the~other sources.
However, some possible improvements are under study~\cite{Raidal:2008jk}, e.g. the~low-$p_\mathrm{T}$ muon identification efficiency could be improved.

\subsection{Expected Reach}

The~predicted upper limits at~$90\%\;\mathrm{C.L.}$ for~the~BR($\tau\to\mu\mu\mu$) from~$Z$ and~$W$ bosons as sources are summarized in~Table~\ref{tab:taumumumuReach}.
The~expected reach in~the~$W$-source channel after the~initial low-luminosity stage of~LHC ($L_\mathrm{int}=30\;\mathrm{fb}^{-1}$) corresponds to~the~present limit set by~the~Belle experiment~\cite{Miyazaki:2007zw}.

\begin{table}[h]
\begin{center}
\caption{Expected reach in~BR($\tau\to\mu\mu\mu$) measurement at~CMS ($90\%\;\mathrm{C.L.}$ limits).}
\begin{tabular}{|l|c|c|}
\hline
$\tau$ lepton source & $L_\mathrm{int}=10\;\mathrm{fb}^{-1}$ & $L_\mathrm{int}=30\;\mathrm{fb}^{-1}$ \\
\hline
$W$ boson decay      & $7.0\cdot 10^{-8}$ & $3.8 \cdot 10^{-8}$ \\
$Z$ boson decay      &         -          & $3.4 \cdot 10^{-7}$ \\
$B$ hadron decay     &         -          & $2.1 \cdot 10^{-7}$ \\
\hline
\end{tabular}
\label{tab:taumumumuReach}
\end{center}
\end{table}

A~significant improvement for~the~contribution from~$B$-mesons is envisaged once a~better performing $b$-tagging algorithm is implemented.
An~improved sensitivity could be reached by~combining results from~all the~sources.
The~measurements will also continue during~the~high luminosity phase of~the~LHC, but~the~studies need to~consider pile-up effects.

\section{\label{sec:Bstaumu} \boldmath$B_s\to\tau\mu$ Particle Level Study}

A~particle level based feasibility study of~the~LFV $B_s\to\tau\mu$ decay has been reported in~\cite{Langenegger:2008zz}.
A~new reconstruction method for~decays with~a~missing particle is used, based on~topological information gained from~secondary vertices~\cite{Dambach:2006ha}.
Given a~3-prong $\tau$ decay and~assuming knowledge of~the~$\tau$ production vertex, the~$\tau$ flight direction is determined as well as the~transverse momenta of~the~3-prong system w.r.t. the~$\tau$ flight direction.
Then the~unknown longitudinal coordinate of~the~neutrino momentum can be calculated with~a~two-fold ambiguity by~constraining the~reconstructed $\tau$ mass.

Given a~$B_s\to\mu\tau(3\;\mathrm{prong})$ decay, the~$\tau$ production vertex is not known, but is required to be located on~the~muon track.
Since the~$B_s$ hadron is expected to~originate from~the~primary vertex, the~$B_s$ decay vertex position is fixed by~a~$B_s$ momentum pointing-to-primary-vertex constraint.

\begin{figure}[h]
\centering
\includegraphics[width=72mm,height=49mm]{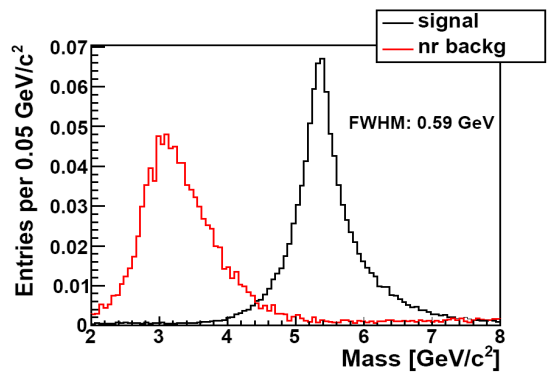}
\caption{Reconstructed mass distribution for~$B_s\to\tau\mu$ and~non-resonant background~\cite{Langenegger:2008zz}. Both histograms are normalized to~unity.}
\label{fig:BstaumuMass}
\end{figure}

In~this~feasibility study, the~MC track parameters and~the~vertex positions were smeared according to~realistic resolutions of~a~general purpose detector for~a~hadron collider experiment, as described in~Table~\ref{tab:BstaumuDetResol}.
The~reconstructed $B_s$ mass distribution has a~non-gaussian shape with~a~FWHM of~$0.59\;\mathrm{GeV}$ as~shown in~Figure~\ref{fig:BstaumuMass}.
A~non-resonant background consisting of~unconstrained $b\bar{b}$ semi-muonic decays with~a~muon and~a~$\tau$ lepton in~the~final state is also presented in~the~Figure~\ref{fig:BstaumuMass} (no normalization to~cross-sections).
Contributions from~exclusive decays as $B\to\mu\nu_\mu D(3\;\mathrm{prong})$ are still to be studied.

\begin{table}[h]
\begin{center}
\caption{MC track and vertex smearing parameters.}
\begin{tabular}{|ll|l|}
\hline
Variable         &                  & Resolution                     \\
\hline
Momentum         & $\phi$           & $0.58\;\mathrm{mrad}$          \\
                 & $\eta$           & $5.8\cdot 10^{-4}$             \\
                 & $1/p_\mathrm{T}$ & $0.013\;\mathrm{(GeV/c)}^{-1}$ \\
\hline
Primary vertex   & xy-plane         & $20\;\mu\mathrm{m}$            \\
                 & z-direction      & $40\;\mu\mathrm{m}$            \\
\hline
Secondary vertex & parallel         & $70\;\mu\mathrm{m}$            \\
                 & perpendicular    & $10\;\mu\mathrm{m}$            \\
\hline
\end{tabular}
\label{tab:BstaumuDetResol}
\end{center}
\end{table}

\vspace*{-5mm}

\vspace*{-5mm}

\section{Summary}

A~set of~ATLAS and~CMS analyses relating lepton flavour violation phenomena to~purely leptonic $B_s$ decays were summarized.
Presently, there is only a~particle based feasibility study of~a~direct LFV measurement on~$B_s \to l_1^+ l_2^-$ decays, but the~phenomena can also be constrained by~measuring the~BR of~the~very rare decay $B_s \to \mu^+ \mu^-$.
In~some models~\cite{Dedes:2002rh} the~BR of~$B_s \to \mu^+\mu^-$ is correlated with~the~BRs of~the~LFV decays $B_s \to l_1^+ l_2^-$ and~$\tau \to \mu \mu \mu$.

The~$B_s \to \mu^+ \mu^-$ studies of~the~decay show that the~BR measurement will improve limits from~Tevatron measurement already after the first year of~LHC running at~low-luminosity.
The~$3\sigma$ measurement of~Standard Model BR is expected by~the~end of~the~low luminosity stage and~$5\sigma$ observation should be made shortly after the~first year of~running at~high luminosity.
A~detailed analysis of~possibly strong background contributions from~other rare decays and particle mis-identification effects was performed, predicting negligible contribution.
Thus the~decays $b\bar{b} \to \mu^+ \mu^- X$ remain the~main source of~background.

The~$90\%\;\mathrm{C.L.}$ upper limits on~the~BR($\tau \to \mu \mu \mu$) measured at~CMS will reach present Belle bounds by~the~end of~LHC's low luminosity stage.
The~analysis of~the~$\tau$ lepton sources predict that the~$\tau$ leptons from~$W$ and~$Z$ boson decays will dominate the~measurements, because the~lower cross-section w.r.t. to~$B$- and~$D$-meson induced $\tau$ decays, is compensated by~a~higher trigger acceptance.

An~initial study of~the~topological reconstruction of~decays with missing particles applied to~the~$B_s \to \tau \mu$ channel demonstrated the~feasibility of~such measurements at~ATLAS and~CMS, but a~proper trigger study, full detector simulation and~a~background analysis still needs to~be performed.

\vspace*{-10mm}

\bigskip 

\end{document}